\documentclass[a4paper,12pt]{article} 
\usepackage{graphicx} 
\usepackage{graphics} 
\newcommand{\dfr}{d\raise0.3ex\hbox{\kern-0.5ex\char"013 }} 

\newcommand{\Csc}{I \!\!\!\! C} %C scale
\newcommand{\Planck}{I \!\!\!\! P} %Planck
\newcommand{\R}{I \!\!R} %rŽels
\begin{document}

\title{Gauge field theory in scale relativity}
\author{L. Nottale, M.N. C\'el\'erier and T. Lehner \\
{\small LUTH, CNRS, Observatoire de Paris-Meudon, 
 F-92195 Meudon Cedex France}}
\maketitle

\abstract
The aim of the present article is to give physical meaning to the 
ingredients of standard gauge field theory in the framework of the scale relativity theory. Owing to the principle of the relativity of scales, the scale-space is 
not absolute. Therefore, the scale variables are functions 
of the space-time coordinates, so that we expect a coupling between the displacement in space-time and the dilation/contraction of the scale variables, which are identified with gauge transformations. The gauge fields naturally appear as a new 
geometric contribution to the total variation of the scale variables. The 
gauge charges emerge as the generators of the scale transformation group 
applied to a generalized action (now identified with the scale relativistic invariant) and are therefore the conservative quantities which find their origin in the 
symmetries of the scale-space. We recover the expression for the covariant 
derivative of non-Abelian gauge theory. Under the gauge 
transformations, the fermion multiplets and the boson field transform in such 
a way that the Lagrangian, which is here derived instead of being set as a 
founding axiom, remains invariant. We have therefore obtained gauge theories 
as a consequence of scale symmetries issued from a geometric fractal 
space-time description, which we apply to peculiar examples of the 
electroweak and grand unified theories. \\

PACS numbers: 11.15.-q, 11.30.-j, 12.10.-g

%1***************
\section{Introduction}
%****************

In standard gauge field theory, the nature of the gauge transformations, 
of the gauge fields and of the conserved charges are not actually understood. 
They are mainly derived from experimental considerations. The group of gauge 
transformations does not act upon the space-time coordinates, as does the 
SU(2) spin rotation group or the Lorentz group, but in an ``internal space'', 
the physical meaning of which is not specified. For a general gauge group G, 
the particle wave functions (represented by $\psi_i$ multiplets of Dirac bi-spinors) form a $n$-component vector in the internal space, 
and the gauge potentials $A_{\mu}$ (more generally $W_{\mu}^i$) are fields in 
standard space-time defined only up to a gauge transformation. 

The aim of the present article is to give physical meaning to the above cited 
different ingredients of the gauge phenomenology in the framework of scale 
relativity. It extends to non-Abelian gauge theory the results of previous 
works \cite{LN94,LN96,LN03} devoted to the understanding of the simpler (Abelian) gauge 
invariant theory of electromagnetism. 

In the theory of scale relativity, space-time is described as a non-differentiable 
``manifold'', which implies that its geometry is fractal \cite{GO83,NS84,LN93}, i.e., explicitly 
depending on the resolution scales \cite{BM82}. Therefore resolutions are not 
only a characteristic of the measurement apparatus, but acquire a 
universal status. They are considered as essential variables, inherent to the 
physical description and characterizing the ``state of scale'' of the 
reference system, in the same way as velocity characterizes its state of 
motion. 

Now, one of the geometric consequences of the fractal character of 
space-time is that there is an infinity of fractal geodesics relating any 
couple of points of this fractal space-time. It is therefore assumed that 
the description of a quantum mechanical system could be reduced to the 
geometric properties of the set of fractal geodesics which corresponds to a 
given state of this system. In such an interpretation, we do not have to 
endow the ``particles'' with internal properties such as mass, spin or charge, 
since they are not considered as point masses which would follow the 
geodesics, but their internal properties are simply defined as geometric 
properties of the geodesics themselves. 

Another consequence of the scale relativistic principles is that the internal 
fractal structures of the geodesic families are smoothed out at scales larger 
than some transition scale  (which is reduced to the Einstein-de Broglie scale for a free particle), related to the 
Compton-length $\lambda$ of the particle under consideration, i.e., to its inertial mass \cite{LN93,LN89}. These 
structures, which are therefore only relevant in the quantum domain, appear at 
scales defined in a relative way. As an example, in the electromagnetic 
theory, only a ratio of scales $\rho={\lambda  / \varepsilon}$ has a physical 
meaning. The relativity of scales, which is, in the scale relativity theory, 
the equivalent of the relativity of motion in Einstein's special and general 
relativity, implies that a displacement of the ``particle'' in 
space-time is linked to a change of the scale of a given structure (of the 
fractal geodesics).
Because the ``scale-space'' is assumed not to be absolute, the 
scale of this structure will not be the same at different locations of the 
bundle of geodesics representing the ``particle''. It is this relative 
``scale-space'' which, in scale relativity, is identified with the ``internal 
space'' of standard gauge theory, as we show below. In non-Abelian gauge 
theory, the scales of the internal fractal structures are characterized 
by a set of scale variables $\eta_a(x,y,z,t)$, generalizing the single 
variable $\rho (x,y,z,t)$ of electromagnetism. 

After a reminder of the results previously obtained for electromagnetism 
(Section 2), we proceed (Section 3) with an extension of the concepts and 
methods thus obtained, and we apply them to a general analysis of the gauge 
formalism. This formalism is, in the new approach, 
related to the coupling between motion in space-time and 
transformations of scale variables occuring in the ``internal'' 
scale space (which is not directly observable). We are therefore able to 
identify gauge transformations with 
scale transformations of the internal ``resolutions", and the charges with the conservative quantities which 
find their origin in the symmetries of the scale-space. 
Moreover, we 
recover the expression for the covariant derivative of non-Abelian gauge 
theory which is therefore no longer postulated, but here derived from geometry and first principles. Section 4 is devoted to the conclusion. 

%2*******************************
\section{Scale relativistic theory of electromagnetism: a reminder}
%********************************

%2.1****************************************
\subsection{ Electromagnetic field and electric charges}
%******************************************

Let us briefly recall the results previously obtained \cite{LN94,LN96,LN03} in the case of a U(1) field (that describes, e.g.,  a scalar spinless charged particle in an electromagnetic field). At scales smaller than the Compton-length $\lambda$ of a particle (in rest frame), the geodesical curves (which constitute an infinite family identified with the ``particle'' itself in the scale relativity framework) are considered to have internal fractal structures. The scales of these structures are defined only in a relative way: namely, only a ratio of scales $\varrho=\lambda / \varepsilon$ does have physical meaning. 

Let us now consider a more general situation in which this scale ratio depends explicitly on the space-time coordinates, i.e., $\varrho=\varrho(x,y,z,t)$. We assume that the scale-space (to which it belongs) remains differentiable. Because the scale-space is considered to be non-absolute (this is another expression of the principle of the relativity of scales), we expect that the scale of a structure will change during a displacement of the particle in space-time. This is analogous to the situation already encountered in general relativity: namely, in a parallel displacement, a vector $V^{\mu}$ is subjected to an increase $\delta V^{\mu}=-\Gamma^{\mu}_{\nu \rho} V^{\nu} dx^{\rho}$ due to the geometric effects of curvature. Then one substracts this geometric increase from its total variation $dV^{\mu}$, in order to recover only the inertial part of the variation (see, e.g., \cite{LL2}, p. 315). This allows one to define the covariant derivative,
\begin{equation}
D V^{\mu}= d V^{\mu} -  \delta V^{\mu}.
\end{equation}
The same kind of behavior is true in the scale relativity framework, but with an essential difference: while the effects of curvature affect only vectors, tensors, etc..., but not scalars, the effects of fractality begins with scalars, among which the ``invariant'' of length $ds^2$ itself. 

\begin{figure}[!ht]
\begin{center}
\includegraphics[width=6cm]{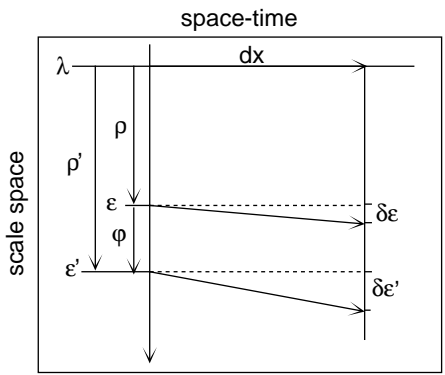}
\caption{\footnotesize{Scale dilations induced by space-time displacements.}}
\label{GFfig1}
\end{center}
\end{figure}

Therefore, in a displacement of the electron we expect the appearance of a resolution change due to the fractal geometry (see Fig. \ref{GFfig1}), that reads
\begin{equation}
\label{0002}
\delta \varepsilon=-\frac{1}{q} \; A_{\mu}\; \varepsilon \; dx^{\mu},
\end{equation}
i.e.,
\begin{equation}
\delta \ln \varrho=\frac{1}{q} \; A_{\mu}\;  dx^{\mu}.
\label{eq3}
\end{equation}

The introduction of the $(1/q)$ term in this definition is an important point for the  electromagnetic case and also for its non-Abelian generalizations, and it must be emphasized from now. Indeed, as we shall see in what follows, the ``field" $A_{\mu}$ will be identified with an electromagnetic potential. Let us take as an example a Coulomb electric potential, $\varphi=q / r$. Since $\ln \varrho$ is dimensionless, we are led to divide the potential term by the ``active'' electric charge $q$, leaving a charge-independent purely geometric contribution.

Let us set
\begin{equation}
\chi=q \ln \varrho.
\end{equation}
This leads to the appearance of a dilation field, according to the construction of a scale-covariant derivative,
\begin{equation}
D \chi= d \chi - \delta \chi= d \chi -   A_{\mu} dx^{\mu}.
\end{equation}
In terms of partial covariant derivative we finally obtain for the sum of the inertial term and of the geometric term:
\begin{equation}
\label{eq2}
\partial_{\mu} \chi= D_{\mu} \chi + A_{\mu}.
\end{equation}

%%%%%%%%%%%%%%%%%%%%%
%%%%%%%%%%%%%%%%%%%%%%

Let us now consider the action $S$ for the electron. In the framework of a space-time theory based on a relativity principle, as it is the case here, it should be given directly by the space-time invariant $s$, i.e., $dS=-mc \, ds$. This relation ensures that the stationary action principle $\delta \int dS=0$ becomes identical with a geodesics (Fermat) principle $\delta \int ds=0$. Now the fractality of the geodesical curves to which the electron wave field is identified means that their proper length is here a function of the scale variable, so that $S= S(\chi)$, at scales smaller than $\lambda$. 

Therefore its differential reads
\begin{equation}
\label{2222}
dS=\frac{\partial S}{\partial \chi} \, d \chi=\frac{\partial S}{\partial \chi} \, (D \chi + A_{\mu} dx^{\mu}),
\end{equation} 
so that we obtain
\begin{equation}
  \partial_{\mu} S =  D_{\mu}S + \frac{\partial S}{\partial \chi} \;  A_{\mu}.
\end{equation} 
This result provides us with a definition for the ``passive'' charge (on which the electromagnetic field acts) \cite{LN94,LN96},
\begin{equation}
\label{5893}
\frac{e}{c} = - \frac{\partial S}{\partial \chi}.
\end{equation} 
This is a second important point worth to be emphasized, since it will play an important role for the generalizations to non-Abelian gauge theories that follow in this paper. In the standard theory, the charge is set from experiment, then it is shown to be  related to gauge transformations, while the gauge functions are considered to be arbitrary and devoid of physical meaning. In the scale relativity approach, the charges are built from the symmetries of the scale space. One indeed recognizes in Eq.(\ref{5893}) the standard expression that relates a conservative quantity to the symmetry of a fundamental variable (here, the internal relative resolution defined in the ``scale-space"), following Noether's theorem. 

We have therefore now established from first principles the form of the action in the classical electromagnetic theory, in particular the form of the particle-field coupling term, which was up to now merely postulated (see e.g. \cite{LL2}),
\begin{equation}
dS= -mc \, ds -\frac{e}{c} \,  A_{\mu}\, dx^{\mu}.
\end{equation} 
But this form has also a new geometric interpretation. It means that an increase of the length can now come from two contributions: the first is the usual variation due to the motion of the particle, and the second new geometric contribution is a length dilation of the internal fractal structure. In other words, the interaction of a charged particle and of an electromagnetic field is, in this framework, described in terms of an energy-momentum transfer between the motion (i.e., the ``external" geometry in general relativity) and the internal geometry.

We are now able to write a geodesics equation minimizing the length-invariant (i.e., the proper time), which identifies with the least-action principle $\delta \int dS=0$ (see \cite{LL2}, p. 85). The variation of the above action yields the Lorentz equation of electrodynamics,
\begin{equation}
mc \, \frac{du_{\alpha}}{ds}= \frac{e}{c}  \, F_{\alpha \mu} \, u^{\mu},
\end{equation}
where $F_{\alpha \mu}$ is the electromagnetic tensor field: $F_{\alpha \mu} = 
\partial_{\alpha}A_{\mu} - \partial_{\mu} A_{\alpha}$. \\
We also recover the standard form for the differential of the action in function of the coordinates, namely,
\begin{equation}
\label{9237}
dS= -(mc \, u_{\mu}+\frac{e}{c}  \, A_{\mu}) \, dx^{\mu}.
\end{equation}

%*****************************
\subsection{Quantum electrodynamics}
%******************************

Let us conclude this reminder by a brief account of the generalization of this approach to quantum electrodynamics. In the scale relativistic quantum description \cite{LN96,LN93,LN03B}, the four-velocity ${\cal V}^{\mu}$ that describes a scalar particle is now complex, so that its action is also a complex number and writes $ S  = S (x^{\mu}, {\cal V}^{\mu}, \eta)$. The wave function is defined from this action as
\begin{equation}
\psi = \exp \left( i \frac{S}{\hbar} \right).
\end{equation}
Therefore Eq.(\ref{9237}) takes the new form,
\begin{equation}
\label{2763}
dS  =   - mc {\cal V} _{\mu} \; dx ^{\mu}  -   \frac{e}{c}  A _{\mu} dx ^{\mu}.    
\end{equation}
The new relation between the wave function and the velocity reads
\begin{equation}
m c {\cal V} _{\mu} = i \hbar  \; D_{\mu}  \ln \psi = i \hbar \partial_{\mu} \ln \psi - \frac{e}{c} A _{\mu} ,
\end{equation}
so that we finally recover the standard QED-covariant derivative as being nothing but the scale-covariant derivative previously introduced, but now acting on the wave function,
\begin{equation}
D _{\mu }  =   \partial  _{\mu } + i \,  \frac{e}{ \hbar c} \, A _{\mu} .   
\end{equation}

%2.3********************
 \subsection{Gauge invariance}
%**********************

Let us now consider another internal structure of the fractal geodesics (identified, in this approach, to the ``particle''), that lies at a relative scale $\varepsilon'=\rho' \lambda$. Equation (\ref{eq3}) becomes
\begin{equation}
\delta \ln \rho' = \frac{1}{q} A'_{\mu} dx^{\mu}. 
\end{equation}
Let $\varphi$ be the ratio between the scales $\varepsilon'$ and $\varepsilon$. In the framework of Galilean scale relativity, this ratio is simply
\begin{equation}
\varphi=\frac{\rho'}{\rho}.
\label{eq18}
\end{equation}

We therefore find
\begin{equation}
A'_{\mu}=A_{\mu} + q\, \partial_{\mu} \ln \varphi.
\end{equation}
In this approach a gauge transformation is identified with a scale transformation of the internal resolutions (in this case, a global dilation). Under such a transformation, the wave function of the particle becomes
\begin{equation}
\psi'= \psi \; \exp \left( -i \frac{e q}{\hbar c} \ln \varphi \right).   
\end{equation}
As a consequence, the full Lagrangian (particle + field-particle coupling) given by Eq.~(\ref{2763}) remains invariant under a gauge transformation.

When $q=e$ (the electron charge), we have $e^2=4 \pi \alpha \hbar c$, where $\alpha$ is the ``fine structure constant'', i.e., the electromagnetic coupling constant. The previous expression becomes
\begin{equation}
\label{2579}
\psi'= \psi \; \exp \left( -i 4 \pi \alpha \ln \varphi \right).  
\end{equation}

%2.3********************
 \subsection{Mass-charge relation}
%**********************

It is worth recalling here, even if this will not be developed further in the present paper, that a log-Lorentzian special theory of scale relativity can also be constructed \cite{LN92}, in which Eq.(\ref{eq18}) becomes
\begin{equation}
\label{9911}
\ln\varphi=\frac{\ln\rho'-\ln\rho}{1-\ln\rho'\ln\rho \, / \, \Csc^2}.
\end{equation}
In this case the possible values of the scale ratios become limited, i.e., $\ln\rho < \Csc, \forall \rho$. The constant $\Csc$ is naturally related to the Planck length-scale $\lambda_{\Planck}=\sqrt{\hbar G/c^3}$, namely, $\Csc=\ln(\lambda/\lambda_{\Planck})=\ln(m_{\Planck}/m)$, where $m_{\Planck}=\sqrt{\hbar c/G}$ is the Planck mass.

While Eq.~(\ref{2579}) leads to no new result in the Galilean scale relativity framework, it implies interesting consequences in the Lorentzian scale relativity framework \cite{LN94,LN96,LN03}. Indeed, in this case $\ln \varphi $ is limited, $\ln\varphi< \Csc$, so that the charge is quantized in terms of a relation that reads (in a spin-like situation where the phase varies of $4 \pi$) 
\begin{equation}
\alpha \; \Csc=1.  
\end{equation}
In other words, one expects a relation between a particle mass-scale and a coupling constant of the form \cite{LN94},
\begin{equation}
\label{8429}
m=m_{\Planck} \times e^{-1/\alpha}.  
\end{equation}
However it would be premature to directly compare this relation to the experimental values of the electron mass and charge. Indeed, it involves a scale ratio that goes up to the Planck scale, while we now know that the low energy electric charge and the electromagnetic field are only large scale residuals of a more complicated electroweak theory, itself issued from a probable chromoelectroweak grand unified theory (GUT). We have suggested \cite{LN94,LN96} that  Eq.~(\ref{8429}) applies to the electron mass using an effective coupling $8 \alpha/3$, where $\alpha$ is the fine structure constant, and where the factor 3/8 originates from GUT \cite{GG74,GQW74}. However, a complete solution must await a full generalization of the scale relativity approach to non-Abelian fields. This is precisely the aim of the present paper to set the bases for such a generalization.

Before generalizing the approach to non-Abelian gauge theories, let us conclude this reminder by noting that it shares some features with the Weyl-Dirac theory of electromagnetism \cite{weyl,dirac}, but with new and essential differences. Namely, the Weyl theory considers scale transformations of the line element, $ds \rightarrow ds'=\rho ds$, but without specifying a fundamental geometric cause for this dilation. The variation of $ds$ should therefore exist at all scales, in contradiction with the observed invariance of the Compton-length of the electron (i.e., of its mass). In the scale relativity proposal, the change of the line element comes from the fractal geometry of space-time, and is therefore a consequence of the dilation of internal resolutions. The new mass-charge relations find their origin in this physical meaning given to the scale transformations. Moreover, the explicit effects of the dependence  on resolutions occur only below the fractal-nonfractal transition, which is identified in rest frame with the Compton scale of the particle. This ensures the invariance of the electron mass in this theory.

%3*********************************
\section{Non-Abelian gauge fields}
%********************************

%3.1****************************
\subsection{Scale relativistic description}
%******************************

%3.1.1****************************
\subsubsection{Introduction}
%******************************

Let us now generalize this description to an attempt of a geometric foundation on first principles of non-Abelian gauge theories. We consider that the internal fractal structures of the ``particle'' (i.e., of the family of geodesics of a non-differentiable space-time) are now described in terms of several scale variables $\eta_{\alpha \beta \ldots} (x,y,z,t)$, that generalize the single resolution variable $\varepsilon$, written for simplicity in $\lambda$ units. We assume that the various indices can be gathered into one common index: we therefore write the scale variables under the simplified form $\eta_{\alpha}$ ($\alpha=0$ to $N$). We still assume them to be differentiable. The precise definition of these variables will be specified in a forthcoming work. In the simplest case, $\eta_{\alpha}=\varepsilon_{\alpha}$, where $\varepsilon_{\alpha}$ is the resolution of the space-time coordinate $X_{\alpha}$ ($\alpha=1$ to $4$). However, more general situations can be considered (5-dimensional penta-velocities of the space-time-djinn \cite{LN03B}, resolution tensor, etc...). In Sec. 4, we give a brief discussion of the simplest case, but we let this point open for more general cases, since our aim is mainly to relate in a general way the scale relativistic tool to the standard description of current gauge theories.

%3.1.2****************************
\subsubsection{General scale transformations}
%******************************

Let us now consider more general scale transformations. In the case of infinitesimal transformations, the transformation law on the $\eta_{\alpha}$ can be written in a linear way:
\begin{equation}
\eta'_{\alpha}= \eta_{\alpha} + \delta \eta_{\alpha}= (\delta_{\alpha \beta} + \delta \theta_{{\alpha}{\beta}}) \, \eta^{\beta},
\end{equation}
where $\delta_{\alpha \beta}$ is the Kronecker symbol, or equivalently,
\begin{equation}
\delta \eta_{\alpha}= \delta \theta_{{\alpha}{\beta}} \; \eta^{\beta}.
\end{equation}
Let us now assume that the $\eta_{\alpha}$'s are functions of the standard space-time coordinates. This leads us to generalize the scale-covariant derivative previously defined in the electromagnetic case as follows: the total variation of the resolution variables becomes the sum of the inertial one, described by the covariant derivative, and of the new geometric contribution, namely,
\begin{equation}
\label{0027}
d \eta_{\alpha}= D \eta_{\alpha} + \eta^{\beta} \: \delta \theta_{{\alpha}{\beta}}=D \eta_{\alpha} + \eta^{\beta} \: W_{{\alpha}{\beta}} ^{\mu} \; dx_{\mu}.
\end{equation}
Recall that in the Abelian case, which corresponds to a unique global dilation, this expression can be simplified since $d\eta/\eta=d \ln \eta =d \chi$.

Therefore, in this new situation we are led to introduce gauge ``fields'' $W_{{\alpha}{\beta}} ^{\mu}$ (more precisely, they will be identified with the gauge potentials) which are linked to the scale transformations as follows,
\begin{equation}
  \delta \theta_{{\alpha}{\beta}}= W_{{\alpha}{\beta}}^{\mu} \; dx_{\mu}.
\end{equation} 
One should remain cautious about this expression and keep in mind that these fields find their origin in a covariant derivative process and are therefore not gradients (this is expressed by the use of a difference sign $\delta \theta_{{\alpha}{\beta}}$ instead of $d \theta_{{\alpha}{\beta}}$). It is also, once again, important to notice that the $W_{{\alpha}{\beta}} ^{\mu}$'s introduced at this level of the analysis do not include charges. They are a function only of the space and time coordinates. This is a necessary choice because our method generates, as we shall see again, not only the fields but also the charges from (respectively) the scale transformations and the scale symmetries of the dynamical fractal space-time. Therefore, as already remarked in previous publications \cite{LN96}, when the scale variables become multiplets, the same is true of the charges. As we shall see in what follows, in the present approach it is at the level of the construction of the charges that the group generators will intervene.

Namely, after having defined the transformation law of the basic variables (the $\eta_{\alpha}$'s), we are now led to describe how various physical quantities transform under the $\eta_{\alpha}$ transformations. These new laws of transformation are expected to depend on the nature of the objects to transform (e.g., vectors, tensors, spinors, etc...), which implies to jump to group representations.

In the case where the particle is a spin 1/2 fermion, it has been shown that the relation between the velocity and the wave function reads \cite{CN01,CN03}
\begin{equation}
{\cal V} _{\mu} = i \lambda \; \psi^{-1} \partial_{\mu}  \psi,   
\end{equation}
where ${\cal V} _{\mu}$ and $\psi$ are complex quaternions (which provides a bijective representation of Dirac bi-spinors). The constant $\lambda=\hbar/mc$ is the Compton length of the particle. Recall that this relation leads to a demonstration of the Dirac equation as an integral of a free-motion-like geodesics equation \cite{CN01}.

However, bispinors are not yet a general enough description for fermions subjected to a general gauge field. Indeed, we consider here a generalized group of transformations (of which the nature will be discussed at the end of this paper), which therefore implies generalized charges. As a consequence of these new charges (that are explicitly defined herebelow), the very nature of the fermions is expected to become more complicated. Experiments have indeed shown that new degrees of freedom must be added in order to represent the weak isospin, hypercharge and color. In order to account in a general way for this more complicated description, we shall simply introduce multiplets $\psi_k$, where each component is a Dirac bi-spinor. In this case the multi-valued velocity becomes a biquaternionic matrix,
\begin{equation}
{\cal V}_{jk} ^{\mu} = i \lambda \;  \psi_j^{-1} \partial^{\mu}  \psi_k.  
\end{equation}
Therefore the action becomes also a two-index quantity,
 \begin{equation}
dS_{jk}=dS_{jk}(x^{\mu},{\cal V}_{jk} ^{\mu},\eta_{\alpha}).
\end{equation}
In the absence of a field, it is now linked to the generalized velocity (and therefore to the wave function) by the relation,
\begin{equation}
\label{5307}
\partial^{\mu}S_{jk}=-m c \; {\cal V}_{jk} ^{\mu}=-i \hbar  \;  \psi_j^{-1} \partial^{\mu}  \psi_k.
\end{equation}

Now, in the presence of a field (i.e., here, when the second-order effects of the fractal geometry are included), using the complete expression for $\partial^{\mu} \eta_{\alpha}$,
\begin{equation}
\partial^{\mu} \eta_{\alpha}=D^{\mu} \eta_{\alpha} + W^{\mu}_{ \alpha \beta} \; \eta^{\beta},
\end{equation}
we are led to write a relation that generalizes Eq.~(\ref{2222}) to the non-Abelian case:
\begin{equation}
\partial^{\mu}S_{jk}=\frac{\partial S_{jk}}{\partial \eta_{\alpha}} \; \partial^{\mu} \eta_{\alpha} =\frac{\partial S_{jk}}{\partial \eta_{\alpha}} \, ( D^{\mu} \eta_{\alpha} + W^{\mu}_{ \alpha \beta} \; \eta^{\beta}).
\end{equation}
Thus we obtain
\begin{equation}
\label{1256}
\partial^{\mu}S_{jk}=D^{\mu} S_{jk}+\eta^{\beta}\; \frac{\partial S_{jk}}{\partial \eta_{\alpha}} \; W^{\mu }_{\alpha \beta}.
\end{equation}

We are finally led to define a general group of scale transformations with generators
\begin{equation}
T^{\alpha \beta}= \eta^{\beta}\partial^{\alpha},
\end{equation}
(where we use the compact notation $\partial^{\alpha}=\partial/ \partial \eta_{\alpha}$), yielding generalized charges
\begin{equation}
\frac{\tilde{g}}{c} \; t^{{\alpha}{\beta}}_{jk}= \eta^{\beta}\; \frac{\partial S_{jk}}{\partial \eta_{\alpha}}.
\end{equation}
Since it must be linked by a unitarity condition (when it is applied on the wave functions, $\psi \psi^{\dagger}$ must be conserved), this is a real special linear group SL(N,$\R$) having $N^2-1$ generators. Such a group is isomorphic to a SU(N) group of complex $N \times N$ matrices (see Sec. 4) \cite{georgi,stancu}.

We have therefore now reached an understanding from first principles, in terms of a geometric space-time description, of the nature of gauge transformations. Let us indeed recall the fundamental difference between the situation of transformations in the standard gauge theories with, e.g., Lorentz transformations. We know from the very beginning what Lorentz transformations are, namely, space-time rotations of the coordinates, i.e., in the case of an infinitesimal transformation,  (i) $dx'^{\alpha}=(1+\omega^{\alpha}_{\beta}) dx^{\beta}$. Then, once this basic definition is given, one can consider the effect of these transformations on various physical quantities $\psi$. This involves the consideration of representations of the Lorentz group adapted to the nature of the physical object under consideration, i.e., (ii) $ \psi'=(1+ \frac{1}{2} \omega^{\alpha \beta} \sigma_ {\alpha \beta}) \psi$ (see, e.g., \cite{SW72}). In the case of the standard theory of gauge transformations, there was up to now no equivalent of the basic defining transformation (i), and the gauge group was directly defined through its action on the various physical objects. It is precisely an equivalent of the defining transformation (i) that we propose in this paper. 

%**************************************
\subsubsection{Rotations in scale-space}
%**************************************

In order to enlight the meaning of the new definition we have obtained for the charges, we shall consider in the present section a sub-sample of the possible scale transformations on internal resolutions: namely, those that can be described in terms of rotations. They constitute the antisymmetric part of the gauge group. In this case the infinitesimal transformation is such that 
\begin{equation}
\delta \theta_{{\alpha}{\beta}}= -\delta \theta_{{\beta}{\alpha}} \Rightarrow W_{{\alpha}{\beta}}^{\mu}=-W_{{\beta}{\alpha}}^{\mu}.
\end{equation}
Therefore, reversing the indices in Eq.(\ref{1256}), we may write
\begin{equation}
\label{4573}
\partial_{\mu}S_{jk}=D_{\mu} S_{jk}+\eta^{\alpha} \; \frac{\partial S_{jk}}{\partial \eta_{\beta}} \;   W_{{\beta}{\alpha}} ^{\mu} .
\end{equation}
Taking the half-sum of Eqs.(\ref{1256}) and (\ref{4573}) we finally obtain
\begin{equation}
\partial_{\mu}S_{jk}=D_{\mu} S_{jk}+ \frac{1}{2}\left( \eta^{\beta}\; \frac{\partial S_{jk}}{\partial \eta_{\alpha}}-\eta^{\alpha} \; \frac{\partial S_{jk}}{\partial \eta_{\beta}}\right) \; W_{{\alpha}{\beta}} ^{\mu} .
\end{equation}

This leads to another definition of the charges,
\begin{equation}
\frac{\tilde{g}}{c} \; t^{{\alpha}{\beta}}_{jk}=-\frac{\partial S_{jk}}{\partial \theta_{{\alpha}{\beta}}} = \frac{1}{2}\left(\eta^{\beta} \; \frac{\partial S_{jk}}{\partial \eta_{\alpha}}    -   \eta^{\alpha}\; \frac{\partial S_{jk}}{\partial \eta_{\beta}}\right).
\end{equation}
We recognize here the form of the definition of an angular momentum from the derivative of the action, i.e., of the conservative quantity that finds its origin in the isotropy of space; but the space under consideration here is the scale-space. Therefore the charges of the gauge fields are identified, in this interpretation, with ``scale-angular momenta''. 

The subgroup of transformations corresponding to these generalized charges is, in three dimensions, a SO(3) group related to a SU(2) group by the homomorphism which associates to two distinct 2$\times$2 unitary matrices of opposite sign the same rotation. We are therefore naturally led to define a ``scale-spin'', which we propose to identify to the simplest non-Abelian charge in the present standard model: the weak isospin. 

Coupling this SU(2) representation of the rotations in scale-space to the U(1) representation of the global scale dilations (that describes the electromagnetism process) analysed in Sec. 2, we are therefore able to give a physical meaning to the transformation group corresponding to the U(1)$\times$SU(2) representation of the standard electroweak theory.

%**********************
\subsubsection{Simplified notation}
%***********************

For the developments to follow, we shall simplify the notation and use only one index $a=(\alpha,\beta)$ for the scale transformations, i.e., running on the gauge group parameters, now written $\theta_a$. For example, in three dimensions this means that  we replace the three rotations $\theta_{23},\theta_{31},\theta_{12}$ respectively by $\theta_{1},\theta_{2},\theta_{3}$. 
We  obtain the following more compact form for the complete action
\begin{equation}
dS_{jk}= \left( D_{\mu} S_{jk}  +  \frac{\tilde{g}}{c} \; t^a_{jk} \; W_{a \mu} \right) dx^{\mu},
\end{equation}
and therefore
\begin{equation}
D^{\mu}S_{jk}=-i \hbar  \;  \psi_j^{-1} D^{\mu}  \psi_k  =-i \hbar  \;  \psi_j^{-1} \partial^{\mu}  \psi_k -\frac{\tilde{g}}{c} \; t^a_{jk} \; W_a^{\mu}.
\end{equation}

%**************************************
\subsection{From scale relativity tools to Yang-Mills theories}
%**************************************

The previous equations used new concepts that are specific of the scale relativity approach, namely the scale variables $\eta_{\alpha}$, the biquaternionic velocity matrix ${\cal V}_{jk} ^{\mu}$ and its associated action $S_{jk}$. The standard basic concepts of quantum field theories, namely the fermionic field $\psi$, the bosonic field $W_a^{\mu}$, the charge $g$, the gauge group generators $t^a_{jk}$ and the gauge-covariant derivative $D_{\mu}$ are now derived from these new concepts.

Let us show that we are able to recover the various relations of standard non-Abelian gauge theories. From the preceding equation, we first recover the standard form for the covariant partial derivative, now acting on the wave function multiplets,
\begin{equation}
D^{\mu} \psi_k= \partial^{\mu} \psi_k  - i \, \frac{\tilde{g}}{\hbar c} \; t^{ja}_{k} \; W_a^{\mu} \;  \psi_j.
\end{equation}
The $\psi_i$'s do not commute together since they are biquaternionic quantities, but this is the case neither of $t^{ja}_{k}$ nor of $W_a^{\mu}$, so that $\psi_j$ can be put to the right as in the standard way of writing; from the multiplet point of view (index $j$), we simply exchange the lines and the columns.

Now introducing a dimensionless coupling constant $\alpha_g$ and a dimensionless charge $g$, such that
\begin{equation}
g^2=4 \pi \alpha_g = \frac{\tilde{g}^2}{\hbar c},
\end{equation}
and redefining the dimensionality of the gauge field (namely, we replace $W_a^{\mu}/\sqrt{\hbar c}$ by $W_a^{\mu}$), the covariant derivative may be more simply written under its standard textbook form,
\begin{equation}
\label{5555}
D^{\mu} \psi_j = \partial^{\mu} \psi_j  - i \, g \; t^{ka}_{j} \; W_a^{\mu} \;  \psi_k.
\end{equation}
In the simplified case of a fermion singlet, it reads
\begin{equation}
D^{\mu}= \partial^{\mu} - i \, g \; t^{a} \; W_a^{\mu} .
\end{equation}

Let us now derive the laws of gauge transformation for the fermion field. Consider a transformation $\theta_a$ of the scale variables. As we shall now see, the $\theta_a$'s can be identified with the standard parameters of a non-Abelian gauge transformation. Indeed, using the above remark about the exchange of lines and columns, Eq.~(\ref{5307}) becomes
\begin{equation}
-i \hbar \partial^{\mu} \psi_j= \partial_{\mu} S^k_j \; \psi _k  .
\end{equation}
and then we obtain Eq. (\ref{5555}), from which we recover the standard form for the transformed fermion multiplet in the case of an infinitesimal gauge transformation $\delta \theta_a$,
\begin{equation}
\psi'_j= ( \delta^k_j - i g \, t^{ka}_j  \, \delta \theta_a) \psi_k  .
\end{equation}
We now have at our disposal all the tools of quantum gauge theories. The subsequent developments are standard ones in terms of these tools. Namely, one introduces the commutator of the matrices $t_a$ (which have a priori no reason to be commutative), \begin{equation}
t_a t_b - t_b t_a= f_{ab}^c \; t_c  .
\end{equation}
Therefore the $t_a$'s are identified with the generators of the gauge group and the $f_{ab}^c$'s with the structure constants of its associated Lie algebra. The non-commutativity finally implies the appearance of an additional term in the boson field law of the gauge transformation. One finds, still in the case of an infinitesimal gauge transformation $\delta \theta^a$, that it transforms according to
\begin{equation}
\label{E26}
\delta W^a _{\mu} = \partial _{\mu} \delta \theta^a + g \; f_{bc}^a \; \delta \theta^b \; W^c _{\mu}  .
\end{equation}
We recognize here once again the standard transformation of non-Abelian gauge theories, which is now derived from the basic transformation Eq.~(\ref{0027}) on the $\eta_a$'s.

Under these gauge transformations, the fermion multiplet and the boson field transform in such a way that the Lagrangian ${\cal L}= \bar{\psi}(i \gamma^{\mu} D_{\mu} -m) \psi$, that contains the fermion and fermion-boson coupling thanks to the covariant derivative remains invariant. It is worth stressing here that, in the scale relativity framework, this form of the Lagrangian (the Dirac form established in \cite{CN01,CN03} and the covariant derivative contribution established here) is derived instead of being set as a founding axiom. The gauge field self-coupling term $-{1\over 4} F_{\mu \nu} F^{\mu \nu}$, which is the simplest gauge invariant scalar that we can add to the Lagrangian, yields the standard Yang-Mills equations. We are therefore provided with a fully consistent gauge theory obtained as a consequence of scale symmetries issued from a geometric space-time description.

%*********************************
\subsection{On the unified gauge group}
%*********************************

We leave to future work a more extensive study of the other 
transformations which might be applied to the scale variables, but we can already consider a more general approach leading to identify a particularly interesting gauge group relevant to the theory. 

In the simplest case of Galilean scale relativity, the scale variables can be 
represented by a four-dimensional ``error" tensor of order two (analogous to a covariance matrix) of the four 
space-time coordinates. At a given point in space-time, we can always choose 
a reference frame where the tensor matrix is diagonal. When we proceed to a displacement 
in space-time from this point, the resolutions $\varepsilon_a$ undergo a 
transformation which, in a first approach limited to the simplest case, we 
can consider as linear in the resolutions. It therefore writes
\begin{equation}
\varepsilon_a = M^b_a \varepsilon_b,
\end{equation}
with $a,b=x,y,z,t$. 

$ M^b_a$ is a $4\times 4$ real matrix, which contains non-diagonal terms 
which are the result of the resolution transformations. Since we are here 
interested in transformations verifying global scale invariance, we limit 
ourselves to the set of matrices with unit determinant. 
The corresponding group is SL(4,$\R$). This group is isomorphic to the SU(4) group which therefore constitutes its appropriate unitary representation applying to the operators 
acting upon Dirac spinor multiplets. 

The above reasoning applies in the simplified framework of the 
four-dimensional Galilean scale relativity. However, a more complete implementation of the principle of scale relativity leads one to introduce a new form of dilation laws [Eq.~(\ref{9911})] having a log-Lorentzian form \cite{LN92}, in which the Planck length-time-scale becomes invariant under dilations and contractions (it therefore plays the role devoted to the zero point in the standard theory). Such a more 
general scale relativistic theory implies to combine the four space-time dimensions with a fifth dimension (that has been called the ``djinn"). This new dimension is related to the fractal dimension of space-time, which is 
no more a constant, but varies explicitly in function of the scale \cite{LN93,LN92}. It is in the framework of such a fifth-dimensional space-time-djinn that the quantization of charge can be established and that new mass-charge relations are obtained (see \cite{LN94,LN96,LN03} and Sec. 2). In this case, the relevant transformation group 
for the resolutions must be fifth-dimensional, which leads us, in the 
simplest linear case, to identify it to the SL(5,$\R$) group. This group is 
isomorphic to SU(5), which becomes the appropriate representation group of the 
operators acting upon the Dirac spinor multiplets. 

We have therefore reached the conclusion that the simplest grand unification gauge group which emerges from a gauge theory developed in the framework of scale 
relativity is SU(5). This result agrees with the first attempts based on group theoretic arguments \cite{GG74} to find a gauge group unifying the U(1) and SU(2) electroweak fields and the SU(3) strong field.

%*********************************
\subsection{Discussion}
%*********************************

However, the group SU(5) was later dismissed as a possible unifying group because the GUT predictions for the weak mixing angle and the proton lifetime were found to contradict experimental results. We recall here that this problem is set in a completely different way in the special scale relativity framework \cite{LN92}, so that it is possible to reconsider working in such a simplest scheme.

The value of the weak mixing angle at unification scale under SU(5) (and some other unification group) is $\sin^2 \theta_w(m_{GUT})=3/8$. By running it down to the W/Z scale using the solutions to the renormalization group equations, one predicts $\sin^2 \theta_w(m_{Z})=0.210$, while recent determinations yield (in the modified minimal subtraction scheme) $\hat{s}^2_Z=0.23113(15)$ \cite{PDG02}.

Another drawback which prevents SU(5) 
from being retained as a relevant gauge group for grand unified theory in the 
standard model is its uncompatibility with the bounds on the proton 
lifetime $t_p$ as they are constrained by, e.g., the Super-Kamiokande data 
which impose $t_p>10^{33}$ yrs \cite{PDG02}. These constraints are 
sufficient to rule out non SUSY or minimal SUSY SU(5) GUT. 

However, in the special scale relativity framework, these questions are set in a fundamentally different way. Indeed, the laws of dilation have a log-lorentzian form (as a direct manifestation of the principle of scale relativity), so that a new relation between length-time-scales and energy-momentum scales is established \cite{LN93,LN92}, that reads (when taking as reference the Z boson scale):
\begin{equation}
\label{108}
\ln \frac{m}{m_{Z}}   = \frac{\ln (\lambda_{Z}/\lambda )}{\sqrt{1 - \frac{\ln ^{2}(\lambda_{Z}/\lambda )}{\ln ^{2}(\lambda_{Z}/\lambda_{\Planck} )}}}    .     
\end{equation}
A major consequence of this new structure of space-time is that the Planck length-time scale becomes invariant under dilations and now plays the role devoted to the zero point. The Planck mass scale is no longer its inverse. From Eq.~(\ref{108}) one finds that it corresponds in the new framework to a length-scale $\lambda_G$ given by $\ln(\lambda_Z/\lambda_G)=\ln(m_{\Planck}/m_Z) / \sqrt{2}$, which is nothing but the unification scale \cite{LN92}. 

Since the effects of gravitation become dominant at that scale, a full unification of the gravitational field with the gauge fields is needed at still larger energy scales. In the scale+motion-relativity framework, gravitation manifests the effects of space-time curvature, and gauge fields manifest the effects of its fractality: when reaching the Planck energy, the curvature has increased so much that it becomes indistinguishable from the fractal fluctuations. The quantum, gravitational and gauge field properties become mixed in a unique extremely complicated geometric behavior, that remains to be understood and to be described (and that cannot be reduced to a quantum gravity theory, since, in the same way as the quantum aspects break the classical gravitational description, the gravitational aspects also break the standard quantum description \cite{LN92}).

Now, when going down to lower energies ($E<E_{\Planck}$), this unique field is spontaneously broken by the rapid decrease of curvature that separates the gravitational field and the gauge fields. In such a scenario, the SU(5) group would be valid only on a small scale-range around the unification scale, and would subsequently be broken in $U(1) \times SU(2) \times SU(3)$ by a cascade effect. One therefore recovers the quantum numbers of hypercharge, isospin and color and the Dirac spinor multiplets corresponding to each of the subgroups in the above direct product. Due to the presence of gravitation at unification scale, one expects the appearance of threshhold effects, so that there is no reason for the three U(1), SU(2) and SU(3) low energy running couplings to converge at exactly the same point. This relaxes the constraint on the weak mixing angle and allows to render its experimental value consistent with the theoretical prediction \cite{LN96}.

As concerns the the proton lifetime, it is given in SU(5) by $t_p\propto m^4_G/(\alpha^2_G m^5_p)$, where $\alpha_G$ is the GUT coupling constant, $m_G$ the GUT mass scale and $m_p$ the proton mass. The standard GUT predicts $m_G \sim 10^{15}$ Gev, which gives $t_p \sim 10^{31}$ yrs. Now, in special scale relativity, $m_G =m_{\Planck}$ \cite{LN96,LN93,LN92}. This adds a factor $(10^4)^4 = 10^{16}$ to the predicted proton lifetime, which therefore becomes compatible with the experimental data. 

It is worth stressing here that the number of degrees of freedom of SU(5) is 
24. We therefore obtain 24 fields of null mass (before symmetry breaking): 12 are the known gauge bosons (8 gluons, 3 SU(2) bosons and one U(1)$_Y$ boson), while the other 12 may have acquired Planck masses in the spontaneous symmetry breaking at the Planck scale (or may correspond to 24 new degrees of freedom allowing for other structures of the theory \cite{LN01}).

%**************************************
\section{Conclusion}
%**************************************

In the present article our purpose has been to give physical meaning to the 
various ingredients of the gauge phenomenology in the framework of scale 
relativity, extending to non-Abelian gauge theory the results of 
previous works \cite{LN94,LN96,LN03} devoted to the understanding of the Abelian gauge-invariant theory of electromagnetism. 

In the theory of scale relativity, space-time is described as a non-differentiable 
``manifold'', which implies that its geometry is fractal, i.e., explicitly 
depending on the resolution scale \cite{LN96,LN93}. Therefore resolutions are 
considered as essential variables, inherent to the 
physical description and characterizing the ``state of scale'' of the 
reference system, in analogy with velocity characterizing its state of 
motion. 

Following the scale relativity principle, we have assumed the scale-space 
not to be absolute, which implies that the $\eta_{\alpha}$'s, which describe the internal dependence on resolutions, are functions 
of the space-time coordinates. Therefore, a displacement in 
space-time is linked to a transformation of the scale variables. In a scale 
transformation, the gauge fields naturally appear in the 
geometric contribution to the total variation of the scale variables. 

In this framework, we have also been led to add the scale variables to the 
usual space-time variables describing the action of the physical system 
(particle). The gauge charges appear as the generators of the group 
of scale transformations applied to this generalized action, therefore 
emerging from the scale symmetries of the dynamical fractal space-time. 

Considering the transformation laws verified by the scale variables, we have 
then been able to establish how the various physical quantities transform 
under these laws and recover the standard gauge theory form of these 
transformations. We have thus established that the fermion multiplet and the 
boson fields transform in such a way that the Lagrangian describing the 
fermion + fermion-boson-coupling remains invariant. 

We are now provided with a theory where the gauge group 
is no more defined through its mere action on the physical objects, as in the 
standard theory, but as the transformation group of the scale variables, and 
where the boson fields and the charges are given their physical meaning 
instead of being put ``by hand". 

Since, in the present study, our aim was to recover from the scale relativistic 
first principles the standard description of current gauge theory, we 
have, in the main part of the work, retained the more general form for 
these scale variables, generically noted $\eta_{\alpha}$. However, besides 
the global scale dilation which yields the classical electromagnetic theory, 
we have considered the non-Abelian case of scale 
rotations which exemplifies the physical nature of the gauge charges in 
this particular choice and leads to a natural physical base for the standard 
weak field theory. We have also considered the other interesting example of 
more general transformations linear in the resolutions. They yield SU(5) as 
the simplest grand unified group consistent with the scale relativistic 
approach. We shall, in a forthcoming work, following its decomposition in subgroups, be more specific  on the new geometric definition of the various charges (hypercharge, weak isospin and color) and analyse them in more detail. The consequences for fermions (including possible generalizations of mass-charge relations) and the question of the renormalization of the theory will also be studied. 

 We shall also be 
led in the future to consider more complicated transformations applying 
to the resolutions and therefore more complicated 
gauge group, with the 
hope that we could thus gain a more profound understanding of the 
mechanisms at work in particle physics. \\

Acknowledgements. We are grateful to Dr. J.E. Campagne for helpful remarks on an earlier version of the manuscript.

%***********
\end{document}